Three Packets of Minerals of the Periodic Table of
Chemical Elements and Chemical Compounds

Mikhail M. Labushev
Department of Ore Deposits Geology and Prospecting, Institute of Mining, Geology and Geotechnology, Siberian Federal University, Russian Federation

**ABSTRACT**

The concepts of α-and β-packets of the periodic table of chemical elements and chemical compounds are defined. The first of the 47 minerals α-packets is composed. In it all minerals are arranged in increasing $I_{av}$ index of proportionality of atomic weights of composing chemical elements, the same way as chemical elements are located in increasing atomic weights in the Periodic table. The packet includes 93 known minerals and two compounds - $N_2O_5$ and $CO_2$ - being actually minerals. B-packet of oxides and hydroxides minerals includes 88 known minerals and five chemical compounds - $N_2O_5$, $CO_2$, CO, $SO_3$ and $SO_2$. Two minerals of the packet have not been determined yet. Besides, β-packet of minerals with sulfur, selenium or arsenic is composed, with one mineral not defined yet. The results of the calculations can be used for further development of the Periodic Table of Chemical Elements and Chemical Compounds and their properties investigation.

**INTRODUCTION**

It is assumed that the number of chemical elements and minerals, as well as inorganic and organic chemical compounds in Nature corresponds to 1, 2, 3 and 4-combinations of a set of 95 and is respectively equal to 95, 4,465, 138,415 and 3,183,545**.** The explanation of these relations is suggested being based on the concept of information coefficient of proportionality $I_p$ as mathematical generalization of proportionality coefficient for any set of positive numbers**.** The indices of proportionality of atomic weights are taken as the mathematical expectations values of $I_{av}$ symmetrized distributions of $I_p$ values [1].

**THE PACKETS OF MINERALS CALCULATION**

The calculations of $I_{av}$ values using PAM method [2] are carried out with the help of Agemarker open source program, available for free calculation at www.skyproject.org. The initial data for the calculations are chemical contents of elements and oxides in nominal mass %.

To calculate $I_{av}$ index, the content of each chemical element is divided into the atomic weight of the respective element and multiplied by the same large number being rounded to inte-ger, then each quotient is multiplied by 8. From this set 8 atomic weights are randomly taken, being then moved into a matrix with three rows and three columns. The ninth element of the matrix is obtained by summing the eight selected atomic weights. The information coefficients of proportionality $I_p$ are calculated for all the elements of the set [1, 2].

For theoretically pure chemical compounds Agemarker calculations can be carried out directly with the numbers of atomic weights of chemical elements. For example, initial data for calculation of ice index can be 2.016 value for hydrogen and 15.999 value for oxygen. These numbers are hereafter divided by the atomic weight of the corresponding element, and 40 million atomic weights of hydrogen and 20 million atomic weights of oxygen, for example, are taken in the calculations. There can be used some other proportionally changed numbers of atomic weights of hydrogen and oxygen, for example 400 and 200 million respectively.

There are considerably more than 95 oxide minerals, with the same number being referred to hydroxides. The selection of minerals of the same packet was made by the following rules:

1. A mineral should not be amorphous.
2. An oxide mineral must contain no more than two chemical elements.
3. A hydroxide mineral must contain no more than three chemical elements.
4. If a mineral except oxygen and hydrogen contains only one chemical element, it refers to the packet of oxides and hydroxides.

There are 88 minerals of this type. Besides, N2O5, CO2, CO, SO3 and SO2 oxides occurring in nature and being able to exist in a crystalline state, were also included into the packet. The new minerals of oxides and hydroxides packet are proposed to be named after the first researchers of periodic changes of chemical elements properties in connection with atomic weights Alexandre-Emile Beguyer de Chancourtois, Julius Lothar Meyer, William Odling, John Newlands and Gustavus Hinrichs: chancourtoisite ($N_2O_5$), meyerite ($CO_2$), odlingite (CO), newlandsite ($SO_3$) and hinrichsite ($SO_2$). The results of the calculations are shown in Table 1.

**Table 1:** $I_{av}$ indexes for minerals of oxides and hydroxides packet

| № | Chemical formula | $I_{av}$ | Standard deviation | Number of $I_p$ calculations, M | Mineral |
|---|---|---|---|---|---|
| 1 | $N_2O_5$ | 0.332406 | 0.009425 | 500 | Chancourtoisite |
| 2 | $CO_2$ | 0.334644 | 0.020820 | 500 | Meyerite |
| 3 | CO | 0.335071 | 0.022840 | 500 | Odlingite |
| 4 | MgO | 0.338624 | 0.033081 | 500 | Periclase |
| 5 | $Al_2O_3$ | 0.342151 | 0.041762 | 500 | Corundum |
| 6 | $SiO_2$ | 0.343090 | 0.044471 | 500 | Cristobalite Coesite Quartz |
| 7 | BeO | 0.344351 | 0.044889 | 500 | Bromellite |
| 8 | $SO_3$ | 0.346614 | 0.052920 | 500 | Newlandsite |

| № | Chemical formula | $I_{av}$ | Standard deviation | Number of $I_p$ calculations, M | Mineral |
|---|---|---|---|---|---|
| 9 | $SO_2$ | 0.348803 | 0.055130 | 500 | Hinrichsite |
| 10 | CaO | 0.361776 | 0.069259 | 500 | Lime |
| 11 | $(Al_2O_3)_5 \cdot H_2O$ | 0.369020 | 0.067487 | 500 | Akdalaite |
| 12 | $TiO_2$ | 0.371583 | 0.086450 | 500 | Akaogiite Anatase Brookite Rutile |
| 13 | $Ti_2O_3$ | 0.373157 | 0.085253 | 500 | Tistarite |
| 14 | $V_2O_5$ | 0.373309 | 0.090911 | 500 | Shcherbinaite |
| 15 | $VO_2$ | 0.375687 | 0.091087 | 500 | Paramontroseite |
| 16 | $V_2VO_5$ | 0.376923 | 0.090356 | 500 | Oxyvanite |
| 17 | $V_2O_3$ | 0.377333 | 0.089582 | 500 | Karelianite |
| 18 | $Cr_2O_3$ | 0.378730 | 0.090988 | 500 | Eskolaite |
| 19 | $Cu_2O$ | 0.379726 | 0.079924 | 500 | Cuprite |
| 20 | $MnO_2$ | 0.380805 | 0.096598 | 500 | Akhtenskite Pyrolusite Ramsdellite |
| 21 | MnO | 0.381569 | 0.088971 | 500 | Manganosite |
| 22 | $Mn_2O_3$ | 0.382530 | 0.094720 | 500 | Bixbyite |
| 23 | FeO | 0.382655 | 0.089922 | 500 | Wüstite |
| 24 | $Fe_2O_3$ | 0.383667 | 0.095809 | 500 | Hematite Maghemite |
| 25 | $Fe_3O_4$ | 0.383795 | 0.094443 | 500 | Magnetite |
| 26 | NiO | 0.385977 | 0.092748 | 500 | Bunsenite |
| 27 | $Cu_2Cu_2O_3$ | 0.387604 | 0.090459 | 500 | Paramelaconite |
| 28 | CuO | 0.391341 | 0.097115 | 500 | Tenorite |
| 29 | ZnO | 0.393268 | 0.098647 | 500 | Zincite |
| 30 | $GeO_2$ | 0.400652 | 0.115779 | 500 | Argutite |
| 31 | $Si_3O_6 \cdot H_2O$ | 0.402364 | 0.090650 | 500 | Silhydrite |
| 32 | $As_2O_3$ | 0.404773 | 0.114155 | 500 | Arsenolite Claudetite |
| 33 | $Fe_{10}O_{14}(OH)_2$ | 0.406592 | 0.108005 | 500 | Ferrihydrite |
| 34 | $SeO_2$ | 0.406764 | 0.121061 | 500 | Downeyite |
| 35 | $MoO_3$ | 0.413036 | 0.135216 | 500 | Molybdite |
| 36 | $ZrO_2$ | 0.417419 | 0.129485 | 500 | Baddeleyite |
| 37 | $HBO_2$ | 0.418176 | 0.099024 | 500 | Clinometaborite |
| 38 | $MoO_2$ | 0.421127 | 0.132318 | 500 | Tugarinovite |
| 39 | PdO | 0.426397 | 0.119814 | 500 | Palladinite |
| 40 | AlO(OH) | 0.428162 | 0.105947 | 500 | Böhmite Diaspore |
| 41 | CdO | 0.429978 | 0.121769 | 500 | Cadmoxite Monteponite |
| 42 | SnO | 0.433497 | 0.123663 | 500 | Romarchite |
| 43 | SnO2 | 0.436390 | 0.143601 | 500 | Cassiterite |
| 44 | SbSbO4 | 0.438165 | 0.144873 | 500 | Clinocervantite |
| 45 | $Sb_2O_3$ | 0.439876 | 0.137776 | 500 | Sénarmontite Valentinite |
| 46 | $TeO_2$ | 0.441400 | 0.147196 | 500 | Paratellurite Tellurite |
| 47 | $CeO_2$ | 0.447737 | 0.151718 | 500 | Cerianite-(Ce) |
| 48 | $V_3O_4(OH)_4$ | 0.452137 | 0.131279 | 500 | Doloresite |
| 49 | CrO(OH) | 0.452205 | 0.131051 | 500 | Bracewellite Grimaldiite Guyanaite |
| 50 | $Be(OH)_2$ | 0.454080 | 0.127962 | 500 | Behoite Clinobehoite |
| 51 | MnO(OH) | 0.454560 | 0.133513 | 500 | Feitknechtite Groutite Manganite |
| 52 | FeO(OH) | 0.455263 | 0.134258 | 500 | Feroxyhyte Goethite Lepidocrocite |
| 53 | CoO(OH) | 0.457595 | 0.136687 | 500 | Heterogenite |
| 54 | $Ta_2O_5$ | 0.459329 | 0.168787 | 500 | Tantite |
| 55 | $B(OH)_3$ | 0.460796 | 0.131486 | 500 | Sassolite |
| 56 | $VVO_2(OH)_3$ | 0.461047 | 0.136635 | 500 | Häggite |
| 57 | HgO | 0.463864 | 0.139328 | 500 | Montroydite |
| 58 | $Mg(OH)_2$ | 0.464782 | 0.131371 | 500 | Brucite |
| 59 | GaO(OH) | 0.464960 | 0.144360 | 500 | Tsumgallite |
| 60 | PbO | 0.465528 | 0.140154 | 500 | Litharge Massicot |

| № | Chemical formula | $I_{av}$ | Standard deviation | Number of $I_p$ calculations, M | Mineral |
|---|---|---|---|---|---|
| 61 | $V_2O_4 \cdot 2H_2O$ | 0.466318 | 0.140399 | 500 | Lenoblite |
| 62 | VO(OH)2 | 0.469651 | 0.142789 | 500 | Duttonite |
| 63 | $Al(OH)_3$ | 0.470961 | 0.137824 | 500 | Bayerite Doyleite Gibbsite Nordstrandite |
| 64 | $PbO_2$ | 0.471819 | 0.168797 | 500 | Plattnerite Scrutinyite |
| 65 | $Pb_2PbO_4$ | 0.471980 | 0.153031 | 500 | Minium |
| 66 | $Tl_2O_3$ | 0.472338 | 0.157569 | 500 | Avicennite |
| 67 | $V_6O_{13} \cdot 8H_2O$ | 0.472464 | 0.145067 | 500 | Vanoxite |
| 68 | $Bi_2O_3$ | 0.473565 | 0.499328 | 500 | Bismite Sphaerobismoite |
| 69 | $Ca(OH)_2$ | 0.477573 | 0.145200 | 500 | Portlandite |
| 70 | $ThO_2$ | 0.477976 | 0.173250 | 500 | Thorianite |
| 71 | $UO_2$ | 0.479302 | 0.174226 | 500 | Uraninite |
| 72 | $(UO_2)_8O_2(OH)_{12} \cdot 10H_2O$ | 0.482057 | 0.181286 | 500 | Metaschoepite |
| 73 | $H_2O$ | 0.482604 | 0.176645 | 500 | Ice |
| 74 | $Fe(OH)_3$ | 0.486588 | 0.158802 | 500 | Bernalite |
| 75 | $Mn(OH)_2$ | 0.486746 | 0.156527 | 500 | Pyrochroite |
| 76 | $UO_4 \cdot 2H_2O$ | 0.486864 | 0.192867 | 500 | Metastudtite |
| 77 | $WO_3 \cdot H_2O$ | 0.486937 | 0.187248 | 500 | Tungstite |
| 78 | | | | 500 | |
| 79 | $Sn_3O_2(OH)_2$ | 0.487554 | 0.160135 | 500 | Hydroromarchite |
| 80 | | | | 500 | |
| 81 | $Ni(OH)_2$ | 0.488687 | 0.159032 | 500 | Theophrastite |
| 82 | $Cu(OH)_2$ | 0.491008 | 0.162074 | 500 | Spertiniite |
| 83 | $Ga(OH)_3$ | 0.491527 | 0.166835 | 500 | Söhngeite |
| 84 | $(UO_2)O_2(H_2O)_2 \cdot 2H_2O$ | 0.491563 | 0.198072 | 500 | Studtite |
| 85 | $UO_2(OH)_2$ | 0.491566 | 0.198061 | 500 | Paulscherrerite |
| 86 | $Zn(OH)_2$ | 0.491818 | 0.163176 | 500 | Ashoverite Sweetite Wülfingite |
| 87 | $WO_2(OH)_2 \cdot H_2O$ | 0.493119 | 0.191232 | 500 | Hydrotungstite |
| 88 | $MoO_3 \cdot 2H_2O$ | 0.494433 | 0.176963 | 500 | Sidwillite |
| 89 | $UO_3 \cdot 2H_2O$ | 0.495015 | 0.200296 | 500 | Paraschoepite |
| 90 | $(UO_2)_8O_2(OH)_{12} \cdot 12H_2O$ | 0.495193 | 0.200340 | 500 | Schoepite |
| 91 | $Cu(OH)_2 \cdot 3H_2O$ | 0.495787 | 0.180577 | 500 | Anthonyite |
| 92 | $Cu(OH)_2 \cdot 2H_2O$ | 0.497209 | 0.179069 | 500 | Calumetite |
| 93 | $U_2(UO_2)_4O_6(OH)_4 \cdot 9H_2O$ | 0.497813 | 0.202956 | 500 | Ianthinite |
| 94 | $In(OH)_3$ | 0.501315 | 0.185886 | 500 | Dzhalindite |
| 95 | BiO(OH) | 0.506587 | 0.191449 | 500 | Daubréeite |

According to the data, oxide and hydroxide minerals can be arranged in the table similar to the Periodic Table (Fig. 1). The minerals density and hardness is changed regularly in the periods and groups of the table. For example, the places of inert gases are occupied by hard and chemically resistant minerals. The density and hardness of known minerals are given in the table according to http://www.webmineral.com/ resource. Unfortunately, there is no information about these physical properties for many of minerals occurring in the form of crystals, visible only under microscope.

| | | | | | | | | | | | | | | | | | |
|---|---|---|---|---|---|---|---|---|---|---|---|---|---|---|---|---|---|
| $N_2O_5$<br>0.332406<br>**2.05**<br>**?** | | | | | | | | | | | | | | | | | $CO_2$<br>0.334644<br>**1.9**<br>**?** |
| $CO$<br>0.335071<br>**1.25**<br>**?** | $MgO$<br>0.338624<br>**3.67-3.9**<br>**6** | | | | | | | | | | | $Al_2O_3$<br>0.342151<br>**4.05**<br>**9** | $SiO_2$<br>0.343090<br>**2.6-2.65**<br>**7** | $BeO$<br>0.344351<br>**3.017**<br>**9** | $SO_3$<br>0.346614<br>**1.92**<br>**?** | $SO_2$<br>0.348803<br>**2.8**<br>**?** | $CaO$<br>0.361776<br>**3.345**<br>**3.5** |
| $(Al_2O_3)_5\cdot H2O$<br>0.369020<br>**3.68**<br>**7** | $TiO_2$<br>0.371583<br>**3.9**<br>**5.5-6** | | | | | | | | | | | $Ti_2O_3$<br>0.373157<br>**?**<br>**8.5** | $V_2O_5$<br>0.373309<br>**3.2-3.36**<br>**3-3.5** | $VO_2$<br>0.375687<br>**4**<br>**?** | $V_2VO_5$<br>0.376923<br>**?**<br>**?** | $V_2O_3$<br>0.377333<br>**4.87**<br>**8-9** | $Cr_2O_3$<br>0.378730<br>**5.18**<br>**8-8.5** |
| $Cu_2O$<br>0.379726<br>**6.1**<br>**3.5-4** | $MnO_2$<br>0.380805<br>**?**<br>**?** | $MnO$<br>0.38156<br>**5.18**<br>**5-6** | $Mn_2O_3$<br>0.382530<br>**4.95**<br>**6-6.5** | $FeO$<br>0.382655<br>**?**<br>**?** | $Fe_2O_3$<br>0.383667<br>**5.3**<br>**6.5** | $Fe_3O_4$<br>0.383795<br>**5.1 - 5.2**<br>**5.5-6** | $NiO$<br>0.385977<br>**6.4-6.8**<br>**5.5** | $Cu_2Cu_2O_3$<br>0.387604<br>**6.106**<br>**4.5** | $CuO$<br>0.391341<br>**6.5**<br>**3.5-4** | $ZnO$<br>0.393268<br>**5.43-5.7**<br>**4-5** | $GeO_2$<br>0.400652<br>**?**<br>**?** | $Si_3O_6\cdot H2O$<br>0.402364<br>**2.141**<br>**1** | $As_2O_3$<br>0.404773<br>**3.7**<br>**1.5** | $Fe_{10}O_{14}(OH)_2$<br>0.406592<br>**3.8**<br>**?** | $SeO_2$<br>0.406764<br>**4.146**<br>**?** | $MoO_3$<br>0.413036<br>**4.72**<br>**3-4** | $ZrO_2$<br>0.417419<br>**5.5-6**<br>**6.5** |
| $HBO_2$<br>0.418176<br>**?**<br>**?** | $MoO_2$<br>0.421127<br>**?**<br>**4.6** | $PdO$<br>0.426397<br>**?**<br>**?** | $AlO(OH)$<br>0.428162<br>**3-3.07**<br>**3** | $CdO$<br>0.429978<br>**?**<br>**?** | $SnO$<br>0.433497<br>**?**<br>**2-2.5** | $SnO_2$<br>0.436390<br>**6.8-7**<br>**6-7** | $SbSbO_4$<br>0.438165<br>**?**<br>**?** | $Sb_2O_3$<br>0.439876<br>**5.2-5.3**<br>**2** | $TeO_2$<br>0.441400<br>**5.6**<br>**1** | $CeO_2$<br>0.447737<br>**7.216**<br>**?** | $V_3O_4(OH)_4$<br>0.452137<br>**3.27-3.33**<br>**?** | $CrO(OH)$<br>0.452205<br>**4.11**<br>**3.5-4.5** | $Be(OH)_2$<br>0.454080<br>**1.91-1.93**<br>**4** | $MnO(OH)$[1]<br>0.454560<br>**4.144**<br>**5.5** | $FeO(OH)$<br>0.455263<br>**4.2**<br>**?** | $CoO(OH)$<br>0.457595<br>**4.3**<br>**3-4** | $Ta_2O_5$<br>0.459329<br>**8.45**<br>**7** |
| $B(OH)_3$<br>0.460796<br>**3.3-3.5**<br>**1** | $VVO_2(OH)_3$<br>0.461047<br>**?**<br>**4.5** | $UO_2$<br>0.479302<br>**6.5 - 10.95**<br>**5-6** | $(UO_2)_8O_2(OH)_{12}\cdot 10H_2O$<br>0.482057<br>**4.69**<br>**2.5** | $H_2O$<br>0.482604<br>**0.9167**<br>**2.5** | $Fe(OH)_3$<br>0.486588<br>**3.32**<br>**4** | $Mn(OH)_2$<br>0.486746<br>**3.26**<br>**2.5-3** | $UO_4\cdot 2H_2O$<br>0.486864<br>**4.67**<br>**?** | $WO_3\cdot H_2O$<br>0.486937<br>**5.5**<br>**2.5** | | $Sn_3O_2(OH)_2$<br>0.487554<br>**?**<br>**?** | | $Ni(OH)_2$<br>0.488687<br>**4**<br>**3.5** | $Cu(OH)_2$<br>0.491008<br>**3.93**<br>**?** | $Ga(OH)_3$<br>0.491527<br>**3.84**<br>**4-4.5** | $(UO_2)O_2(H_2O)_2\cdot 2H_2O$<br>0.491563<br>**3.64**<br>**1-2** | $UO_2(OH)_2$<br>0.491566<br>**?**<br>**?** | $Zn(OH)_2$[5]<br>0.491818<br>**3.05**<br>**3** |
| $WO_2(OH)_2\cdot H_2O$<br>0.493119<br>**4.6**<br>**2** | $MoO_3\cdot 2H_2O$<br>0.494433<br>**3.12**<br>**2** | | | | | | | | | | | | | | | | |
| | | | | $HgO$<br>0.463864<br>**11.2**<br>**1.5-2** | $Mg(OH)_2$<br>0.464782<br>**2.39-2.4**<br>**2.5-3** | $GaO(OH)$<br>0.464960<br>**?**<br>**1.5-2.5** | $PbO$<br>0.465528<br>**9.14-9.355**<br>**2** | $V_2O_4\cdot 2H_2O$<br>0.466318<br>**?**<br>**?** | $VO(OH)_2$<br>0.469651<br>**3.24**<br>**2.5** | $Al(OH)_3$[2]<br>0.470961<br>**2.3 - 2.4**<br>**2.5-3** | $PbO_2$[3]<br>0.471819<br>**8.5 - 9.63**<br>**5.5** | $Pb_2PbO_4$<br>0.471980<br>**8.2**<br>**2.5-3** | $Tl_2O_3$<br>0.472338<br>**9.574**<br>**2** | $V_6O_{13}\cdot 8H_2O$<br>0.472464<br>**?**<br>**?** | $Bi_2O_3$[4]<br>0.473565<br>**8.5 - 9.5**<br>**4-5** | $Ca(OH)_2$<br>0.477573<br>**2.23**<br>**2.5-3** | $ThO_2$<br>0.477976<br>**10**<br>**6** |
| $HgO$<br>0.463864<br>**11.2**<br>**1.5-2** | ----Formula<br>----$I_{av}$ value<br>**----Density**<br>**--Hardness** | | | $UO_3\cdot 2H_2O$<br>0.495015<br>**?**<br>**2-3** | $(UO_2)_8O_2(OH)_{12}\cdot 12H_2O$<br>0.495193<br>**4.8** 2.5 | $Cu(OH)_2\cdot 3H_2O$<br>0.495787<br>**?**<br>**2** | $Cu(OH)_2\cdot 2H_2O$<br>0.497209<br>**?**<br>**2** | $U_2(UO_2)_4O_6(OH)_4\cdot 9H_2O$<br>0.497813<br>**5.16**<br>**2** | $In(OH)_3$<br>0.501315<br>**4.34**<br>**4-4.5** | $BiO(OH)$<br>0.506587<br>**6.5**<br>**2-2.5** | | | | | | | |

**Figure 1:** Periodic Table of the oxide and hydroxide minerals (β-packet of minerals)

[1] Groutite [2]Gibbsite [3]Plattnerite [4]Bismite [5]Wülfingite

The variation interval of $I_{av}$ indices of minerals in the packet partially coincides with the same intervals in other packets of minerals. These packets are proposed to be named as β-packets of minerals. In contrast, $I_{av}$ indices of minerals in α-packet do not form any crossing intervals.

The total number of packets of chemical compounds is 35,015, where minerals comprise 47, inorganic chemicals 1457 and organic chemical compounds 33,511 packets. The following numbering is proposed: according to it chemical elements and minerals constitute packets with numbers varying from αA1 to αA47 (αA1: αA47); inorganic chemical compounds make up packets αB1: αB1457; organic compounds are characterized by packets α(C1:C1457), α(D1: D1457), ..., α(Y1 :Y1457). The serial number of chemical element or chemical compound in the packet is placed before the number of the packet, for example α1A1 or α95A1**.** The packet of oxide and hydroxide minerals can be characterized as β(A1:A47). Subsequently, it will be possible to reduce the uncertainty of its position relative to α-packet of minerals, such as β(A1:A24) or β(A1:A7).

B-packet minerals with sulfur, selenium or arsenic include all known minerals that contain no more than two chemical elements (Fig. 2). The results of the calculations are shown in Table 2.

**Table 2**: $I_{av}$ indexes for minerals with sulfur, selenium or arsenic of the β-packet

| № | Chemical formula | $I_{av}$ | Standard deviation | Number of $I_p$ calculations, M | Mineral |
|---|---|---|---|---|---|
| 1 | $ZnAs_2$ | 0.332488 | 0.010117 | 500 | Noname, haven't been found in Nature yet |
| 2 | $Cu_3As$ | 0.332665 | 0.011758 | 500 | Domeykite |
| 3 | $Cu_5As_2$ | 0.332735 | 0.012214 | 500 | Koutekite |
| 4 | $CuAs_2$ | 0.332783 | 0.012181 | 500 | Paxite |
| 5 | $MoSe_2$ | 0.333201 | 0.015048 | 500 | Drysdallite |
| 6 | ZnSe | 0.333250 | 0.015089 | 500 | Stilleite |
| 7 | $CuSe_2$ | 0.333467 | 0.015941 | 500 | Krutaite |
| 8 | $CoAs_3$ | 0.333495 | 0.015905 | 500 | Skutterudite |
| 9 | $Cu_2Se$ | 0.333530 | 0.016793 | 500 | Bellidoite |
| 10 | $Cu_3Se_2$ | 0.333650 | 0.017268 | 500 | Umangite |
| 11 | $Cu_5Se_4$ | 0.333692 | 0.017392 | 500 | Athabascaite |
| 12 | CaS | 0.333808 | 0.017828 | 500 | Oldhamite |
| 13 | $CoAs_2$ | 0.333820 | 0.017558 | 500 | Clinosafflorite |
| 14 | $NiAs_2$ | 0.333888 | 0.017845 | 500 | Krutovite |
| 15 | CoAs | 0.334112 | 0.019160 | 500 | Langisite |
| 16 | $Ni_{11}As_8$ | 0.334155 | 0.019508 | 500 | Maucherite |
| 17 | NiAs | 0.334191 | 0.019482 | 500 | Nickeline |
| 18 | $CoSe_2$ | 0.334755 | 0.021222 | 500 | Trogtalite |
| 19 | $FeAs_2$ | 0.334783 | 0.021315 | 500 | Loullingite |
| 20 | $NiSe_2$ | 0.334836 | 0.021505 | 500 | Kullerudite |
| 21 | MgS | 0.334856 | 0.022071 | 500 | Niningerite |
| 22 | $PdSe_2$ | 0.335033 | 0.023235 | 500 | Verbeekite |
| 23 | $RuAs_2$ | 0.335053 | 0.023306 | 500 | Anduoite |
| 24 | $Ag_2Se$ | 0.335144 | 0.022563 | 500 | Naumannite |

| № | Chemical formula | $I_{av}$ | Standard deviation | Number of $I_p$ calculations, M | Mineral |
|---|---|---|---|---|---|
| 25 | $Ni_3Se_4$ | 0.335189 | 0.023027 | 500 | Troustedtite |
| 26 | CoSe | 0.335202 | 0.023300 | 500 | Freboldite |
| 27 | FeAs | 0.335235 | 0.023403 | 500 | Westerveldite |
| 28 | $Rh_2As$ | 0.335257 | 0.022934 | 500 | Rhodarsenide |
| 29 | NiSe | 0.335296 | 0.023621 | 500 | Makinenite |
| 30 | $Pd_{17}Se_{15}$ | 0.335309 | 0.023561 | 500 | Palladseite |
| 31 | $Rh_{12}As_7$ | 0.335439 | 0.023650 | 500 | Polkanovite |
| 32 | $Pd_8As_3$ | 0.335508 | 0.023485 | 500 | Arsenopalladinite |
| 33 | RhAs | 0.335789 | 0.025248 | 500 | Cherepanovite |
| 34 | $FeSe_2$ | 0.335888 | 0.024910 | 500 | Dzharkenite |
| 35 | $Pd_2As$ | 0.335995 | 0.025221 | 500 | Palladoarsenide |
| 36 | FeSe | 0.336533 | 0.027514 | 500 | Achavalite |
| 37 | CdSe | 0.336716 | 0.028049 | 500 | Cadmoselite |
| 38 | $VS_4$ | 0.337544 | 0.032405 | 500 | Patronite |
| 39 | $Cr_3S_4$ | 0.340815 | 0.038647 | 500 | Brezinaite |
| 40 | SbAs | 0.340912 | 0.038241 | 500 | Stibarsen |
| 41 | $MnS_2$ | 0.342159 | 0.042526 | 500 | Hauerite |
| 42 | $FeS_2$ | 0.342787 | 0.043849 | 500 | Pyrite Marcasite |
| 43 | MnS | 0.342916 | 0.042246 | 500 | Alabandite |
| 44 | $FeFe_2S_4$ | 0.343569 | 0.044269 | 500 | Greigite |
| 45 | $Fe_7S_8$ | 0.343633 | 0.043951 | 500 | Pyrrhotite |
| 46 | $Ni_3S_2$ | 0.344714 | 0.044324 | 500 | Heazlewoodite |
| 47 | $NiS_2$ | 0.344788 | 0.047860 | 500 | Vaesite |
| 48 | $CoS_2$ | 0.344962 | 0.048190 | 500 | Cattierite |
| 49 | NiS | 0.345663 | 0.047167 | 500 | Millerite |
| 50 | $Co_9S_8$ | 0.345664 | 0.046795 | 500 | Cobaltpentlandite |
| 51 | $NiNi_2S_4$ | 0.345684 | 0.048155 | 500 | Polydymite |
| 52 | CoS | 0.345841 | 0.047471 | 500 | Jaipurite |
| 53 | $CoCo_2S_4$ | 0.345860 | 0.048471 | 500 | Linnaeite |
| 54 | $Cu_2S$ | 0.346517 | 0.046249 | 500 | Chalcocite |
| 55 | $As_4S$ | 0.346628 | 0.044028 | 500 | Duranusite |
| 56 | $Cu_{31}S_{16}$ | 0.346703 | 0.046631 | 500 | Djurleite |
| 57 | $Cu_9S_5$ | 0.347121 | 0.047507 | 500 | Digenite |
| 58 | $Cu_{1.78}S$ | 0.347183 | 0.047635 | 500 | Roxbyite |
| 59 | $Cu_7S_4$ | 0.347269 | 0.047834 | 500 | Anilite |
| 60 | Cu8S5 | 0.347733 | 0.048838 | 500 | Geerite |
| 61 | $CuS_2$ | 0.348297 | 0.054255 | 500 | Villamannite |
| 62 | $Cu_{39}S_{28}$ | 0.348353 | 0.050272 | 500 | Spionkopite |
| 63 | $Cu_{1.2}S$ | 0.348879 | 0.051627 | 500 | Yarrowite |
| 64 | CuS | 0.349290 | 0.052974 | 500 | Covellite |
| 65 | ZnS | 0.350671 | 0.055020 | 500 | Sphalerite |
| 66 | $As_4S_3$ | 0.356560 | 0.061333 | 500 | Dimorphite |
| 67 | AsS | 0.357823 | 0.064564 | 500 | Realgar |
| 68 | $As_2S_3$ | 0.357878 | 0.067174 | 500 | Orpiment |
| 69 | $As_8S_9$ | 0.358056 | 0.065579 | 500 | Alacranite |
| 70 | $As_4S_5$ | 0.358113 | 0.066318 | 500 | Uzonite |
| 71 | $PtSe_2$ | 0.359818 | 0.071751 | 500 | Sudovikovite |
| 72 | $Pt_5Se_4$ | 0.359931 | 0.065646 | 500 | Luberoite |
| 73 | $OsAs_2$ | 0.361410 | 0.073881 | 500 | Omeiite |
| 74 | $IrAs_2$ | 0.362015 | 0.074692 | 500 | Iridarsenite |
| 75 | HgSe | 0.362607 | 0.070192 | 500 | Tiemannite |
| 76 | PbSe | 0.364520 | 0.072329 | 500 | Clausthalite |

| № | Chemical formula | $I_{av}$ | Standard deviation | Number of $I_p$ calculations, M | Mineral |
|---|---|---|---|---|---|
| 77 | $Bi_2Se_3$ | 0.365293 | 0.076434 | 500 | Guanajuatite |
| 78 | $Ag_2S$ | 0.371149 | 0.073091 | 500 | Acanthite |
| 79 | $RuS_2$ | 0.375010 | 0.090352 | 500 | Laurite |
| 80 | $Rh_{17}S_{15}$ | 0.376202 | 0.082824 | 500 | Miassite |
| 81 | $Rh_2S_3$ | 0.377881 | 0.090139 | 500 | Bowieite |
| 82 | $Rh_3S_4$ | 0.378050 | 0.088971 | 500 | Kingstonite |
| 83 | CdS | 0.382949 | 0.090175 | 500 | Greenockite |
| 84 | $SnS_2$ | 0.386040 | 0.101972 | 500 | Berndtite |
| 85 | SnS | 0.386603 | 0.093262 | 500 | Herzenbergite |
| 86 | $Sn_2S_3$ | 0.387825 | 0.099691 | 500 | Ottemannite |
| 87 | $Tl_2S$ | 0.403974 | 0.094977 | 500 | Carlinite |
| 88 | $WS_2$ | 0.417836 | 0.437534 | 500 | Tungstenite |
| 89 | $Bi_4S_3$ | 0.418285 | 0.109447 | 500 | Ikunolite |
| 90 | $ReS_2$ | 0.418773 | 0.130525 | 500 | Rheniite |
| 91 | PtS | 0.420454 | 0.116505 | 500 | Braggite |
| 92 | HgS | 0.422337 | 0.117558 | 500 | Cinnabar |
| 93 | $Ir_2S_3$ | 0.422932 | 0.126916 | 500 | Kashinite |
| 94 | PbS | 0.424487 | 0.118767 | 500 | Galena |
| 95 | $Bi_2S_3$ | 0.428972 | 0.130865 | 500 | Bismuthinite |

| | | | | | | | | | | | | | | | | | |
|---|---|---|---|---|---|---|---|---|---|---|---|---|---|---|---|---|---|
| $ZnAs_2$?<br>0.332488<br>**?**<br>**?** | | | | | | | | | | | | | | | | | $Cu_3As$<br>0.332665<br>**7.2 - 8.1**<br>**3-3.5** |
| $Cu_5As_2$<br>0.332735<br>**8.48**<br>**3.5** | $CuAs_2$<br>0.332783<br>**5.3**<br>**3.5-4** | | | | | | | | | | | $MoSe_2$<br>0.333201<br>**6.25**<br>**1-1.5** | $ZnSe$<br>0.333250<br>**5.42**<br>**?** | $CuSe_2$<br>0.333467<br>**?**<br>**?** | $CoAs_{3-x}$<br>0.333495<br>**6.1 - 6.9**<br>**?** | $Cu_2Se$<br>0.333530<br>**7.03**<br>**1.5-2** | $Cu_3Se_2$<br>0.333650<br>**5.62 -6.78**<br>**3** |
| $Cu_5Se_4$<br>0.333692<br>**?**<br>**2.5-3** | $CaS$<br>0.333808<br>**2.58**<br>**4** | | | | | | | | | | | $CoAs_2$<br>0.333820<br>**7.46**<br>**4.5-5** | $NiAs_2$<br>0.333888<br>**?**<br>**5.5** | $CoAs$<br>0.334112<br>**8.17**<br>**?** | $Ni_{11}As_8$<br>0.334155<br>**7.83**<br>**5** | $NiAs$<br>0.334191<br>**7.78 - 7.8**<br>**5.5** | $CoSe_2$<br>0.334755<br>**?**<br>**7** |
| $FeAs_2$<br>0.334783<br>**?**<br>**?** | $NiSe_2$<br>0.334836<br>**6.72**<br>**5.5-6.5** | $MgS$<br>0.334856<br>**7.1 - 7.7**<br>**5** | $PdSe_2$<br>0.335033<br>**?**<br>**5.5** | $RuAs_2$<br>0.335053<br>**4.95**<br>**6-6.5** | $Ag_2Se$<br>0.335144<br>**6.5 - 8**<br>**2.5** | $Ni_3Se_4$<br>0.335189<br>**?**<br>**2.5** | $CoSe$<br>0.335202<br>**?**<br>**2.5-3** | $FeAs$<br>0.335235<br>**?**<br>**5.5-6** | $Rh_2As$<br>0.335257<br>**11.27-11.32**<br>**?** | $NiSe$<br>0.335296<br>**?**<br>**2.5-3** | $Pd_{17}Se_{15}$<br>0.335309<br>**8.3**<br>**?** | $Rh_{12}As_7$<br>0.335439<br>**10.22**<br>**?** | $Pd_8As_3$<br>0.335508<br>**10.4**<br>**4** | $RhAs$<br>0.335789<br>**9.72**<br>**?** | $FeSe_2$<br>0.335888<br>**7.34**<br>**5** | $Pd_2As$<br>0.335995<br>**10.42**<br>**?** | $FeSe$<br>0.336533<br>**6.53**<br>**2.5** |
| $CdSe$<br>0.336716<br>**5.663**<br>**4** | $VS_4$<br>0.337544<br>**2.82**<br>**2** | $Cr_3S_4$<br>0.340815<br>**4.12**<br>**3.5-4.5** | $SbAs$<br>0.340912<br>**6.1 - 6.2**<br>**3-4** | $MnS_2$<br>0.34215<br>**3.463**<br>**4** | $FeS_2$<br>0.34278<br>**5 - 5.02**<br>**6.5** | $MnS$<br>0.34291<br>**3.95-4.04**<br>**?** | $FeFe_2S_4$<br>0.34357<br>**4.049**<br>**4-4.5** | $Fe_7S_8$<br>0.343633<br>**4.58- 4.65**<br>**3.5-4** | $Ni_3S_2$<br>0.344714<br>**5.82**<br>**4** | $NiS_2$<br>0.344788<br>**4.45**<br>**4.5-5.5** | $CoS_2$<br>0.344962<br>**4.8**<br>**4.5** | $NiS$<br>0.345663<br>**5.5**<br>**3-3.5** | $Co_9S_8$<br>0.345664<br>**?**<br>**4 - 5** | $NiNi_2S_4$<br>0.345684<br>**4.5 - 4.8**<br>**4.5-5.5** | $CoS$<br>0.345841<br>**5.45**<br>**?** | $CoCo_2S_4$<br>0.345860<br>**4.8**<br>**?** | $Cu_2S$<br>0.346517<br>**5.5 - 5.8**<br>**?** |
| $As_4S$<br>0.346628<br>**4.5**<br>**2** | $Cu_{31}S_{16}$<br>0.346703<br>**5.63**<br>**2.5-3** | $Cu_9S_5$<br>0.347121<br>**5.6**<br>**2.5-3** | $Pt_5Se_4$<br>0.359931<br>**13.02**<br>**5-5.5** | $OsAs_2$<br>0.361410<br>**11.2**<br>**?** | $IrAs_2$<br>0.362015<br>**10.9**<br>**5-5.5** | $HgSe$<br>0.362607<br>**8.19-8.47**<br>**?** | $PbSe$<br>0.364520<br>**7.6 - 8.8**<br>**?** | $Bi_2Se_3$<br>0.365293<br>**6.25-6.98**<br>**2.5-3.5** | $Ag_2S$<br>0.371149<br>**7.2 - 7.4**<br>**2-2.5** | $RuS_2$<br>0.375010<br>**6.99**<br>**?** | $Rh_{17}S_{15}$<br>0.376202<br>**?**<br>**?** | $Rh_2S_3$<br>0.377881<br>**?**<br>**7** | $Rh_3S_4$<br>0.378050<br>**?**<br>**?** | $CdS$<br>0.382949<br>**3.98 - 5**<br>**3.5-4** | $SnS_2$<br>0.386040<br>**4.5**<br>**1-2** | $SnS$<br>0.386603<br>**5.197**<br>**2** | $Sn_2S_3$<br>0.387825<br>**4.835**<br>**2** |
| $Tl_2S$<br>0.403974<br>**8.1**<br>**1** | $WS_2$<br>0.417836<br>**7.4**<br>**2.5** | | | | | | | | | | | | | | | | |
| | | | | $Cu_{1.78}S$<br>0.347183<br>**5.555**<br>**2.5-3** | $Cu_7S_4$<br>0.347269<br>**5.68**<br>**3-4** | $Cu_8S_5$<br>0.347733<br>**?**<br>**3.5-4** | $CuS_2$<br>0.348297<br>**4.523**<br>**4.5** | $Cu_{39}S_{28}$<br>0.348353<br>**5.13**<br>**2.5-3** | $Cu_{1.2}S$<br>0.348879<br>**4.89**<br>**3.5** | $CuS$<br>0.349290<br>**4.6 - 4.76**<br>**?** | $ZnS$<br>0.350671<br>**3.9 - 4.2**<br>**?** | $As_4S_3$<br>0.356560<br>**3.58 - 3.6**<br>**?** | $AsS$<br>0.357823<br>**3.56**<br>**1.5-2** | $As_2S_3$<br>0.357878<br>**3.49-3.56**<br>**1.5-2** | $As_8S_9$<br>0.358056<br>**3.4 - 3.46**<br>**1.5** | $As_4S_5$<br>0.358113<br>**3.37**<br>**1.5** | $PtSe_2$<br>0.359818<br>**9.63 - 9.7**<br>**2-2.5** |
| $Cu_3As$<br>0.332665<br>**7.2 - 8.1**<br>**3-3.5** | ----Formula<br>----$I_{av}$ value<br>**----Density**<br>**--Hardness** | | | $Bi_4S_3$<br>0.418285<br>**7.8**<br>**2** | $ReS_2$<br>0.418773<br>**?**<br>**?** | $PtS$<br>0.420454<br>**10**<br>**1.5** | $HgS$<br>0.422337<br>**8.1**<br>**2-2.5** | $Ir2S_3$<br>0.422932<br>**9.1**<br>**?** | $PbS$<br>0.424487<br>**7.2 - 7.6**<br>**?** | $Bi_2S_3$<br>0.428972<br>**6.8 - 7.2**<br>**?** | | | | | | | |

**Figure 2:** Periodic Table of the minerals with sulfur, selenium or arsenic (β-packet of minerals)

The basis of αA1 packet is comprised by natural alloys, minerals with sulfur, arsenic, selenium, antimony or tellurium (Table 3, Figure 3).

**Table 3:** $I_{av}$ indexes for natural alloys, minerals with sulfur, arsenic, selenium, antimony or tellurium and others of the αA1 packet

| № | Chemical formula | $I_{av}$ | Standard deviation | Number of $I_p$ calculations, M | Mineral |
|---|---|---|---|---|---|
| 1 | $CuZn_2$ | 0.331870 | 0.002143 | 500 | Danbaite |
| 2 | CuZn | 0.331874 | 0.002281 | 500 | Zhanghengite |
| 3 | $HgPb_2$ | 0.331879 | 0.002441 | 500 | Leadamalgam |
| 4 | (Ni,Fe) | 0.331905 | 0.003225 | 500 | Taenite |
| 5 | $Ni_3Fe$ | 0.331915 | 0.003423 | 500 | Awaruite |
| 6 | $AuPb_3$ | 0.331918 | 0.003484 | 500 | Novodneprite |
| 7 | $Fe_{1.5}Cr_{0.2}$ | 0.331925 | 0.003617 | 500 | Chromferide |
| 8 | $Cr_{1.5}Fe_{0.2}$ | 0.331927 | 0.003762 | 500 | Ferchromide |
| 9 | $AuPb_2$ | 0.331932 | 0.003804 | 500 | Anyuiite |
| 10 | Au2Pb | 0.331933 | 0.003851 | 500 | Hunchunite |
| 11 | FeNi | 0.331940 | 0.003988 | 500 | Tetrataenite |
| 12 | CoFe | 0.331957 | 0.004314 | 500 | Wairauite |
| 13 | $Au_2Bi$ | 0.331966 | 0.004506 | 500 | Maldonite |
| 14 | $PtBi_2$ | 0.332008 | 0.005156 | 500 | Insizwaite |
| 15 | $Pd_3Sn$ | 0.332202 | 0.007734 | 500 | Atokite |
| 16 | $Ag_{3.2}Sb_{0.8}$ | 0.332221 | 0.007970 | 500 | Dyscrasite |
| 17 | KCl | 0.332223 | 0.007848 | 500 | Sylvite |
| 18 | $Pd_2Sn$ | 0.332267 | 0.008363 | 500 | Paolovite |
| 19 | $Pd_3Sn_2$ | 0.332300 | 0.008645 | 500 | Stannopalladinite |
| 20 | $KCaCl_3$ | 0.332318 | 0.008831 | 500 | Chlorocalcite |
| 21 | $N_2O_5$ | 0.332406 | 0.009425 | 500 | Chancourtoisite |
| 22 | $ZnAs_2$ | 0.332488 | 0.010117 | 500 | Noname, haven't been found in Nature yet |
| 23 | $Cu_3As$ | 0.332665 | 0.011758 | 500 | Domeykite Domeykite-β |
| 24 | $Cu_5As_2$ | 0.332735 | 0.012215 | 500 | Koutekite |
| 25 | $CuAs_2$ | 0.332783 | 0.012180 | 500 | Paxite |
| 26 | $(Pd,Ag)_3Te$ | 0.332835 | 0.012937 | 500 | Telargpalite |
| 27 | $Pd_{20}Te_7$ | 0.332866 | 0.013135 | 500 | Keithconnite |
| 28 | AgI | 0.332889 | 0.013003 | 500 | Iodargyrite |
| 29 | $Ag_{5-x}Te_3$ | 0.332914 | 0.013280 | 500 | Stützite |
| 30 | $Pd_3SbTe_4$ | 0.332965 | 0.013364 | 500 | Borovskite |
| 31 | $Ag_4Pd_3Te_4$ | 0.332967 | 0.013629 | 500 | Sopcheite |
| 32 | $MoSe_2$ | 0.333201 | 0.015048 | 500 | Drysdallite |
| 33 | ZnSe | 0.333250 | 0.015089 | 500 | Stilleite |
| 34 | NaF | 0.333281 | 0.015246 | 500 | Villiaumite |
| 35 | $CuSe_2$ | 0.333467 | 0.015940 | 500 | Krutaite |
| 36 | $CoAs_{3-x}$ | 0.333494 | 0.015906 | 500 | Skutterudite |
| 37 | $Cu_2Se$ | 0.333529 | 0.016794 | 500 | Bellidoite |
| 38 | $Cu_3Se_2$ | 0.333651 | 0.017267 | 500 | Umangite |
| 39 | $Cu_{5.2}Se_6$ | 0.333687 | 0.017208 | 500 | Klochmannite |
| 40 | $Cu_5Se_4$ | 0.333693 | 0.017392 | 500 | Athabascaite |
| 41 | $NaMgF_3$ | 0.333723 | 0.017667 | 500 | Neighborite |
| 42 | CaS | 0.333808 | 0.017827 | 500 | Oldhamite |
| 43 | $CoAs_2$ | 0.333819 | 0.017558 | 500 | Clinosafflorite Safflorite |
| 44 | $NiAs_2$ | 0.333887 | 0.017845 | 500 | Krutovite Pararammelsbergite Rammelsbergite |

| № | Chemical formula | $I_{av}$ | Standard deviation | Number of $I_p$ calcula-tions, M | Mineral |
|---|---|---|---|---|---|
| 45 | $MgF_2$ | 0.334014 | 0.019097 | 500 | Sellaite |
| 46 | $Ag_8Te_3Se$ | 0.334101 | 0.018637 | 500 | Kurilite |
| 47 | CoAs | 0.334114 | 0.019158 | 500 | Langisite Modderite |
| 48 | $Ni_{11}As_8$ | 0.334156 | 0.019508 | 500 | Maucherite |
| 49 | NiAs | 0.334192 | 0.019482 | 500 | Nickeline |
| 50 | $Na_3AlF_6$ | 0.334220 | 0.020111 | 500 | Cryolite |
| 51 | Pd11Te2Se2 | 0.334280 | 0.019397 | 500 | Miessiite |
| 52 | K4MnCl6 | 0.334396 | 0.021702 | 500 | Chlormanganokalite |
| 53 | NiAsSe | 0.334418 | 0.020034 | 500 | Jolliffeite |
| 54 | FeNi2As2 | 0.334492 | 0.020914 | 500 | Oregonite |
| 55 | $Pd_{11}Sb_2As_2$ | 0.334506 | 0.019867 | 500 | Isomertierite |
| 56 | $Na_5Al_3F_{14}$ | 0.334561 | 0.021611 | 500 | Chiolite |
| 57 | $CO_2$ | 0.334644 | 0.020820 | 500 | Meyerite |
| 58 | $(Ni,Co,Cu)Se_2$ | 0.334683 | 0.020945 | 500 | Penroseite |
| 59 | $CoSe_2$ | 0.334755 | 0.021223 | 500 | Trogtalite |
| 60 | $FeAs_2$ | 0.334781 | 0.021316 | 500 | Löllingite |
| 61 | $NiSe_2$ | 0.334836 | 0.021505 | 500 | Kullerudite |
| 62 | MgS | 0.334857 | 0.022072 | 500 | Niningerite |
| 63 | $Pd_{11}As_2Te_2$ | 0.334876 | 0.021411 | 500 | Törnroosite |
| 64 | $PdSe_2$ | 0.335034 | 0.023235 | 500 | Verbeekite |
| 65 | $(Cu,Ag)_{21}As_{10}$ | 0.335050 | 0.024413 | 500 | Novákite |
| 66 | $RuAs_2$ | 0.335051 | 0.023308 | 500 | Anduoite |
| 67 | $Ag_2Se$ | 0.335146 | 0.022563 | 500 | Naumannite |
| 68 | $Ni_3Se_4$ | 0.335189 | 0.023027 | 500 | Trüstedtite Wilkmanite |
| 69 | CoSe | 0.335203 | 0.023300 | 500 | Freboldite |
| 70 | FeAs | 0.335234 | 0.023404 | 500 | Westerveldite |
| 71 | $CuFeSe_2$ | 0.335250 | 0.023266 | 500 | Eskebornite |
| 72 | $Rh_2As$ | 0.335256 | 0.022933 | 500 | Rhodarsenide |
| 73 | $Pd_3As$ | 0.335295 | 0.022711 | 500 | Vincentite |
| 74 | NiSe | 0.335297 | 0.023619 | 500 | Sederholmite Mäkinenite |
| 75 | $Pd_{17}Se_{15}$ | 0.335309 | 0.023560 | 500 | Palladseite |
| 76 | AgBr | 0.335377 | 0.023892 | 500 | Bromargyrite |
| 77 | $Ag_2Pd_3Se_4$ | 0.335394 | 0.023767 | 500 | Chrisstanleyite |
| 78 | $Rh_{12}As_7$ | 0.335439 | 0.023650 | 500 | Polkanovite |
| 79 | $Sb_2(Sb,As)_2$ | 0.335473 | 0.022420 | 500 | Paradocrasite |
| 80 | $Pd_8As_3$ | 0.335510 | 0.023485 | 500 | Stillwaterite Arsenopalladinite |
| 81 | PdAsSe | 0.335684 | 0.025501 | 500 | Kalungaite |
| 82 | $K_3VS_4$ | 0.335745 | 0.026054 | 500 | Colimaite |
| 83 | RhAs | 0.335790 | 0.025249 | 500 | Cherepanovite |
| 84 | $(Pd,Cu)_7Se_3$ | 0.335853 | 0.024664 | 500 | Oosterboschite |
| 85 | $FeSe_2$ | 0.335890 | 0.024909 | 500 | Dzharkenite Ferroselite |
| 86 | $Pd_2As$ | 0.335994 | 0.025218 | 500 | Palladoarsenide Palladodymite |
| 87 | $(Fe,Co)As_3$ | 0.336079 | 0.024269 | 500 | Ferroskutterudite |
| 88 | $NaNO_3$ | 0.336370 | 0.028317 | 500 | Nitratine |
| 89 | FeSe | 0.336534 | 0.027513 | 500 | Achavalite |
| 90 | $MgCl_2$ | 0.336620 | 0.027004 | 500 | Chloromagnesite |
| 91 | CdSe | 0.336717 | 0.028048 | 500 | Cadmoselite |
| 92 | (Ir,Os,Ru) | 0.336748 | 0.025374 | 500 | Rutheniridosmine |
| 93 | PdSbSe | 0.336846 | 0.027982 | 500 | Milotaite |
| 94 | (Ru,Ni)As | 0.337159 | 0.029301 | 500 | Ruthenarsenite |
| 95 | $AgAuTe_4$ | 0.337463 | 0.032002 | 500 | Sylvanite |

| | | | | | | | | | | | | | | | | | |
|---|---|---|---|---|---|---|---|---|---|---|---|---|---|---|---|---|---|
| $CuZn_2$<br>0.331870<br>**7.36**<br>**4** | | | | | | | | | | | | | | | | | $CuZn$<br>0.331874<br>**?**<br>**3.5** |
| $HgPb_2$<br>0.331879<br>**?**<br>**1.5** | (Ni,Fe)<br>0.331905<br>**7.8 - 8.22**<br>**?** | | | | | | | | | | | $Ni_3Fe$<br>0.331915<br>**8**<br>**5** | $AuPb_3$<br>0.331918<br>**?**<br>**?** | $Fe_{1.5}Cr_{0.2}$<br>0.331925<br>**?**<br>**4** | $Cr_{1.5}Fe_{0.2}$<br>0.331927<br>**?**<br>**6.5** | $AuPb_2$<br>0.331932<br>**?**<br>**3.5** | $Au_2Pb$<br>0.331933<br>**16**<br>**3.5** |
| $FeNi$<br>0.331940<br>**8.275**<br>**?** | $CoFe$<br>0.331957<br>**?**<br>**5** | | | | | | | | | | | $Au_2Bi$<br>0.331966<br>**15.46**<br>**1.5-2** | $PtBi_2$<br>0.332008<br>**12.8**<br>**?** | $Pd_3Sn$<br>0.332202<br>**14.9**<br>**5** | $Ag_{3.2}Sb_{0.8}$<br>0.332221<br>**9.4 - 10**<br>**?** | $KCl$<br>0.332223<br>**1.99**<br>**2.5** | $Pd_2Sn$<br>0.332267<br>**11.32**<br>**5** |
| $Pd_3Sn_2$<br>0.332300<br>**10.2**<br>**5** | $KCaCl_3$<br>0.332318<br>**?**<br>**2.5-3** | $N_2O_5$<br>0.332406<br>**?**<br>**?** | ZnAs2?<br>0.332488<br>**?**<br>**?** | $Cu_3As$<br>0.332665<br>**7.2 - 8.1**<br>**3-3.5** | $Cu_5As_2$<br>0.332735<br>**8.48**<br>**3.5** | $CuAs_2$<br>0.332781<br>**5.3**<br>**3.5-4** | $(Pd,Ag)_3Te$<br>0.33283<br>**7.375**<br>**2- 2.5** | $Pd_{20}Te_7$<br>0.332867<br>**?**<br>**?** | $AgI$<br>0.332887<br>**5.5 - 5.7**<br>**1.5-2** | $Ag_{5-x}Te_3$<br>0.332913<br>**8**<br>**?** | $Pd_3SbTe_4$<br>0.332964<br>**8.12**<br>**3** | $Ag_4Pd_3Te_4$<br>0.332967<br>**?**<br>**3.5** | MoSe2<br>0.333199<br>**6.25**<br>**1-1.5** | $ZnSe$<br>0.333254<br>**5.42**<br>**?** | $NaF$<br>0.333281<br>**2.79 - 2.8**<br>**?** | CuSe2<br>0.333467<br>**?**<br>**?** | $CoAs_{3-x}$<br>0.333495<br>**6.1 - 6.9**<br>**?** |
| $Cu_2Se$<br>0.333532<br>**7.03**<br>**1.5-2** | $Cu_3Se_2$<br>0.333650<br>**5.62-6.78**<br>**3** | $Cu_5Se_4$<br>0.333693<br>**?**<br>**2.5-3** | $NaMgF_3$<br>0.333724<br>**3.03**<br>**4.5** | $CaS$<br>0.333809<br>**2.58**<br>**4** | $CoAs_2$<br>0.333819<br>**7.46**<br>**4.5-5** | $NiAs_2$<br>0.333889<br>**?**<br>**5.5** | $MgF_2$<br>0.334012<br>**2.97-3.15**<br>**5-6** | $Ag_8Te_3Se$<br>0.334104<br>**?**<br>**?** | $CoAs$<br>0.334117<br>**8.17**<br>**?** | $Ni_{11}As_8$<br>0.334158<br>**7.83**<br>**5** | $NiAs$<br>0.334187<br>**7.78 - 7.8**<br>**5.5** | $Na_3AlF_6$<br>0.334222<br>**2.95 - 3**<br>**2.5-3** | $Pd_{11}Te_2Se_2$<br>0.334279<br>**?**<br>**2-2.5** | $K_4MnCl_6$<br>0.334398<br>**2.31**<br>**2.5** | $NiAsSe$<br>0.334420<br>**7.1**<br>**6-6.5** | $FeNi_2As_2$<br>0.334494<br>**6.92**<br>**5** | $Pd_{11}Sb_2As_2$<br>0.334505<br>**?**<br>**?** |
| $Na_5Al_3F_{14}$<br>0.334560<br>**2.84 - 2.9**<br>**3.5-4** | $CO_2$<br>0.334644<br>**1.9**<br>**?** | $CoSe_2$<br>0.334757<br>**?**<br>**7** | $Rh_2As$<br>0.335255<br>**11.27-11.32**<br>**?** | $Pd_3As$<br>0.335293<br>**6.816**<br>**5** | $NiSe$<br>0.335299<br>**?**<br>**2.5-3** | $Pd_{17}Se_{15}$<br>0.335304<br>**8.3**<br>**?** | $AgBr$<br>0.335378<br>**5.8 - 6**<br>**1.5-2** | $Ag_2Pd_3Se_4$<br>0.335393<br>**8.31**<br>**5** | $Rh_{12}As_7$<br>0.335438<br>**10.22**<br>**?** | $Sb_2(Sb,As)_2$<br>0.335473<br>**6.52**<br>**?** | $Pd_8As_3$<br>0.335505<br>**10.4**<br>**4** | $PdAsSe$<br>0.335684<br>**?**<br>**?** | $K_3VS_4$<br>0.335750<br>**?**<br>**?** | $RhAs$<br>0.335783<br>**9.72**<br>**?** | $(Pd,Cu)_7Se_5$<br>0.335852<br>**8.48**<br>**5** | $FeSe_2$<br>0.335884<br>**7.34**<br>**5** | $Pd_2As$<br>0.335993<br>**10.42**<br>**?** |
| $(Fe,Co)As_3$<br>0.336075<br>**?**<br>**?** | $NaNO_3$<br>0.336367<br>**2.24-2.29**<br>**?** | | | | | | | | | | | | | | | | |
| | | | | $(Ni,Co,Cu)Se_2$<br>0.334684<br>**6.58-6.74**<br>**?** | $FeAs_2$<br>0.334784<br>**7.1 - 7.7**<br>**5** | $NiSe_2$<br>0.334836<br>**6.72**<br>**5.5-6.5** | $MgS$<br>0.334856<br>**?**<br>**?** | $Pd_{11}As_2Te_2$<br>0.334878<br>**?**<br>**?** | $PdSe_2$<br>0.335035<br>**?**<br>**5.5** | $(Cu,Ag)_{21}As_{10}$<br>0.335048<br>**6.7**<br>**3-3.5** | $RuAs_2$<br>0.335056<br>**?**<br>**?** | $CO$<br>0.335071<br>**1.25**<br>**?** | $Ag_2Se$<br>0.335148<br>**6.5 - 8**<br>**2.5** | $Ni_3Se_4$<br>0.335190<br>**?**<br>**?** | $CoSe$<br>0.335208<br>**?**<br>**2.5-3** | $FeAs$<br>0.335226<br>**?**<br>**5.5-6** | $CuFeSe_2$<br>0.335249<br>**5.35**<br>**?** |
| $CuZn_2$<br>0.331870<br>**7.36**<br>**4** | ----Formula<br>----$I_{av}$ value<br>**----Density**<br>**--Hardness** | | | $FeSe$<br>0.336528<br>**6.53**<br>**2.5** | $MgCl_2$<br>0.336625<br>**2.325**<br>**?** | $CdSe$<br>0.336712<br>**5.663**<br>**4** | (Ir,Os,Ru)<br>0.336744<br>**?**<br>**6-7** | $PdSbSe$<br>0.336847<br>**7.95-8.23**<br>**?** | (Ru,Ni)As<br>0.337167<br>**10**<br>**?** | $AgAuTe_4$<br>0.337459<br>**7.9-8.3**<br>**1.5-2** | | | | | | | |

Figure 3: αA1 packet of the natural alloys, minerals with sulfur, arsenic, selenium, antimony or tellurium and others

There are no identified specific differences between the minerals of αA1 packet and the minerals of αA2 packet (Table 4), typical for the minerals of the two characterized β packets.

**Table 4:** Some minerals of αA2 packet

| № | Chemical formula | $I_{av}$ | Standard deviation | Number of $I_p$ calculations, mln | Mineral |
|---|---|---|---|---|---|
| 1 | $VS_4$ | 0.337546 | 0.032403 | 500 | Patronite |
| 2 | $Cu_2Pd_3Se_4$ | 0.337707 | 0.031008 | 500 | Jagüeite |
| 3 | (Ag,Hg)I | 0.337841 | 0.033552 | 500 | Tocornalite |
| 4 | $FeCl_3$ | 0.338198 | 0.033784 | 500 | Molysite |
| 5 | $Ag_3CuSe_2$ | 0.338267 | 0.031718 | 500 | Selenojalpaite |
| 6 | $(Au,Ag)_{1.2}Hg_{0.8}$ | 0.338334 | 0.029625 | 500 | Weishanite |
| 7 | $Au_3AgTe_8$ | 0.338428 | 0.034159 | 500 | Krennerite |
| 8 | $Cu_3SbSe_4$ | 0.338447 | 0.034774 | 500 | Permingeatite |
| 9 | (Ag,Cu)I | 0.338459 | 0.030316 | 500 | Miersite |
| 10 | $(Au,Sb)_2Te_3$ | 0.338482 | 0.034098 | 500 | Montbrayite |
| 11 | $Pt_3Sn$ | 0.338528 | 0.031111 | 500 | Rustenburgite |
| 12 | MgO | 0.338627 | 0.033081 | 500 | Periclase |
| 13 | $MnCl_2$ | 0.338703 | 0.034424 | 500 | Scacchite |

In the 19th century chemists were discussing the similarity of chemical elements and saturated hydrocarbons [3]. N.A. Morozov tried to express the analogy in the form of the table with eight vertical rows - classes of hydrocarbons (Table 5).

**Table 5:** The System of Hydrocarbons According to N.A. Morozov

| **Type** | | | | | | | |
|---|---|---|---|---|---|---|---|
| 7 | 6 | 5 | 4 | 3 | 2 | 1 | 0 |
| | | | C | CH | $CH_2$ | $CH_3$ | **$CH_4$** |
| | $C_2$ | $C_2H$ | $C_2H_2$ | $C_2H_3$ | $C_2H_4$ | $C_2H_5$ | **$C_2H_6$** |
| $C_3H$ | $C_3H_2$ | $C_3H_3$ | $C_3H_4$ | $C_3H_5$ | $C_3H_6$ | $C_3H_7$ | **$C_3H_8$** |
| $C_4H_3$ | $C_4H_4$ | $C_4H_5$ | $C_4H_6$ | $C_4H_7$ | $C_4H_8$ | $C_4H_9$ | **$C_4H_{10}$** |
| $C_5H_5$ | $C_5H_6$ | $C_5H_7$ | $C_5H_8$ | $C_5H_9$ | $C_5H_{10}$ | $C_5H_{11}$ | **$C_5H_{12}$** |
| $C_6H_7$ | $C_6H_8$ | $C_6H_9$ | $C_6H_{10}$ | $C_6H_{11}$ | $C_6H_{12}$ | $C_6H_{13}$ | **$C_6H_{14}$** |
| $C_7H_9$ | $C_7H_{10}$ | $C_7H_{11}$ | $C_7H_{12}$ | $C_7H_{13}$ | $C_7H_{14}$ | $C_7H_{15}$ | **$C_7H_{16}$** |
| $C_8H_{11}$ | $C_8H_{12}$ | $C_8H_{13}$ | $C_8H_{14}$ | $C_8H_{15}$ | $C_8H_{16}$ | $C_8H_{17}$ | **$C_8H_{18}$** |
| $C_9H_{13}$ | $C_9H_{14}$ | $C_9H_{15}$ | $C_9H_{16}$ | $C_9H_{17}$ | $C_9H_{18}$ | $C_9H_{19}$ | **$C_9H_{20}$** |
| $C_{10}H_{15}$ | $C_{10}H_{16}$ | $C_{10}H_{17}$ | $C_{10}H_{18}$ | $C_{10}H_{19}$ | $C_{10}H_{20}$ | $C_{10}H_{21}$ | **$C_{10}H_{22}$** |
| $C_{11}H_{17}$ | $C_{11}H_{18}$ | $C_{11}H_{19}$ | $C_{11}H_{20}$ | $C_{11}H_{21}$ | $C_{11}H_{22}$ | $C_{11}H_{23}$ | **$C_{11}H_{24}$** |

Note. In Tables 5 and 6 the bold shows chemically inactive compounds according to N.A. Morozov.

According to N.A. Morozov in the table (representing a shorter version of the periodic table) hydrocarbons are located in columns the same way as chemical elements are arranged in the periodic table by D.I. Mendeleev. N.A. Morozov pointed out that in all groups in his table, except zero, there are chemically active substances. On this basis, he predicted the existence of a new group of chemical elements being later discovered and described as inert gases [4].

The places of inert gases in the periodic table of hydrocarbons arranged in ascending $I_{av}$ index are occupied by chemical compounds $C_2H$ $C_5H_{12}$ $C_{11}H_{24}$ $C_7H_{10}$. Therefore, this table isn't considered to be the only version of displaying regular variations of chemical compounds properties.

Table 6 The part of β-packet of hydrocarbons arranged in ascending values of $I_{av}$ index

| | | | | | | | | | |
|---|---|---|---|---|---|---|---|---|---|
| $C_3H$<br>0.41064 | | | | | | | | | $C_2H$<br>0.43150 |
| $\mathbf{CH_4}$<br>0.44425 | $C_3H_2$<br>0.44563 | $C_4H_3$<br>0.45069 | $CH_3$<br>$\mathbf{C_2H_6}$<br>0.45646 | $\mathbf{C_3H_8}$<br>0.46028 | CH<br>…<br>$C_5H_5$<br>0.46119 | $C_2H_5$<br>$\mathbf{C_4H_{10}}$<br>0.46235 | | | $\mathbf{C_5H_{12}}$<br>0.46347 |
| $C_3H_7$<br>$\mathbf{C_6H_{14}}$<br>0.46438 | $\mathbf{C_7H_{16}}$<br>0.46451 | $C_4H_9$<br>$\mathbf{C_8H_{18}}$<br>0.46509 | $C_6H_7$<br>0.46517 | $\mathbf{C_9H_{20}}$<br>0.46518 | $\mathbf{C_{10}H_{22}}$<br>$C_5H_{11}$<br>0.46531 | $C_5H_6$<br>0.46550 | | | $\mathbf{C_{11}H_{24}}$<br>0.46559 |
| $C_6H_{13}$<br>0.46569 | $C_7H_{15}$<br>0.46596 | $C_8H_{17}$<br>0.46603 | $C_{10}H_{21}$<br>0.46620 | $C_9H_{19}$<br>0.46623 | $C_4H_5$<br>0.46632 | $C_{11}H_{23}$<br>0.46637 | $C_7H_9$<br>0.46680 | $CH_2$<br>…<br>$C_8H_{16}$<br>0.46706 | $C_3H_4$<br>$C_6H_8$<br>0.46737 |
| $C_{11}H_{21}$<br>0.46760 | $C_{10}H_{19}$<br>0.46763 | $C_8H_{11}$<br>0.46764 | $C_9H_{17}$<br>0.46774 | $C_7H_{13}$<br>0.46780 | $C_8H_{15}$<br>0.46784 | $C_5H_7$<br>0.46789 | | | $C_7H_{10}$<br>0.46800 |
| $C_6H_{11}$<br>0.46808 | $C_5H_9$<br>$C_{10}H_{18}$<br>0.46808 | $C_9H_{13}$<br>0.46811 | $C_{11}H_{20}$<br>0.46811 | $C_4H_7$<br>0.46825 | $C_9H_{16}$<br>0.46829 | $C_8H_{14}$<br>0.46835 | $C_2H_3$<br>…<br>$C_{10}H_{15}$<br>0.46841 | $C_5H_8$<br>0.46845 | $C_{10}H_{17}$<br>0.46845 |
| $C_9H_{14}$<br>0.46846 | $C_7H_{12}$<br>0.46846 | $C_7H_{11}$<br>0.46849 | $C_3H_5$<br>$C_9H_{15}$<br>0.46854 | $C_{11}H_{17}$<br>0.46855 | $C_{11}H_{18}$<br>0.46855 | $C_8H_{13}$<br>0.46863 | | | |

To study the packets of chemical compounds it is proposed to use some alternative arrangement of chemical elements in the periodic table.

## CONCLUSION

The Periodic table of chemical compounds packets calculations are based on the atomic weights of chemical elements that are fundamental physical quantities. Thus the Periodic Table of chemical compounds has physical rather than chemical basis.

It is proposed to make all publishing on the periodicity of chemical compounds as open as possible because the results of the systematization of chemical compounds in α-and β-packets can be used in different fields of study, medical and environmental research including.

**ACKNOWLEDGEMENTS**

Author thanks Helen Fomina for the assistance in the translation of the article into English and Timothy Labushev for the joint development and support of Agemarker program.

**REFERENCES**

1**.** Labushev, M. M. (2011). The Periodic Table as a Part of the Periodic Table of Chemical Compounds, 18. Retrieved from http://arxiv.org/abs/1103.3972

2. Labushev, M. M., Khokhlov A.N. (2012). Relative Dating and Classification of Minerals and Rocks Based on Statistical Calculations Related to Their Potential Energy Index, 19. Retrieved from http://arxiv.org/abs/1212.2628

3. Babaev, E.V., Hefferlin, R. (1996). The Concepts of Periodicity and Hyper-Periodicity: from Atoms to Molecules. In: Concepts in Chemistry: a Contemporary Challenge. (Ed. D.Rouvray). Research Studies Press, London, pp.24–81.

4. Morozov, N.A. (1907). Periodicheskie sistemy stroeniya veschestva: Teoriya obrazo-vaniya khimicheskikh elementov.– M.: Sytin,. XVI, 437 s.